\documentclass[twocolumn,showpacs,preprintnumbers,amsmath,amssymb,prx,nofootinbib,aps]{revtex4-2}

\usepackage[pagewise]{lineno}

\usepackage{graphicx}
\usepackage{dcolumn}
\usepackage{bm}
\usepackage{xcolor}
\usepackage{rotating}
\usepackage[normalem]{ulem}
\graphicspath{{figures/}}

\begin{document}
\newcommand\blfootnote[1]{%
	\begingroup
	\renewcommand\thefootnote{}\footnote{#1}%
	\addtocounter{footnote}{-1}%
	\endgroup}
    
\newcommand{\sh}[1]{\textcolor{red}{#1}}

 \newcommand{\FigCap}[1]{\textbf{#1}}	
\newcommand{\blue}{\textcolor{blue}}
\newcommand{\green}{\textcolor{green}}
\newcommand{\mh}{\textcolor{blue}}
\newcommand{\red}{\textcolor{red}}

\newcommand{\ml}{\textcolor{magenta}}

\title{A continuous confinement–deconfinement transition in a triangular quantum magnet}

\setlength{\textfloatsep}{10pt plus 1pt minus 2pt}

\author{Suguru Hosoi$^{1,*}$}
\author{Sejun Park$^1$}
\author{Michihiro Hirata$^1$}
\author{Minseong Lee$^2$}
\author{Adam P Dioguardi$^1$}
\author{Joe D Thompson$^1$}
\author{Filip Ronning$^1$}
\author{Allen O Scheie$^1$}
\author{Kumpei Imamura$^{3,4}$}
\author{Ken-ichiro Hashimoto$^4$}
\author{Takasada Shibauchi$^3$}
\author{Bishnu P Belbase$^5$}
\author{Arjun Unnikrishnan$^{5,6,7}$}
\author{Johannes Knolle$^{8, 9, 10}$}
\author{Arnab Banerjee$^5$}
\author{Yuji Matsuda$^{1,\dagger}$}

{\let\thefootnote\relax\footnote{corresponding authors: \\
$^{*}$shosoi@lanl.gov, ~~$^{\dagger}$matsuda@lanl.gov}}

\affiliation{$^1$Materials Physics and Applications $-$ Quantum, Los Alamos National Laboratory, Los Alamos, New Mexico  87545, USA}
\affiliation{$^2$National High Magnetic Field Laboratory, Los Alamos National Laboratory, Los Alamos, New Mexico  87545, USA}
\affiliation{$^3$Department of Advanced Materials Science, University of Tokyo, Kashiwa, Chiba 277-8561, Japan}
\affiliation{$^4$Department of Physics, Kyoto University, Kyoto 606-8504, Japan}
\affiliation{$^5$Department of Physics and Astronomy, Purdue University, West Lafayette, Indiana 47906, USA} 
\affiliation{$^6$Experimental Physics VI, Center for Electronic Correlations and Magnetism, Institute of Physics, University of Augsburg, 86159 Augsburg, Germany.}
\affiliation{$^7$Solid State and Structural Chemistry Unit, Indian Institute of Science, Bengaluru 560012, India}
\affiliation{$^8$Technical University of Munich, TUM School of Natural Sciences, Physics Department, Garching, Germany}
\affiliation{$^9$Munich Center for Quantum Science and Technology (MCQST), Schellingstr. 4, 80799 M\"{u}nchen, Germany}
\affiliation{$^{10}$Blackett Laboratory, Imperial College London, London SW7 2AZ, United Kingdom}

\begin{abstract}
A continuous transition between phases hosting distinct excitations---bosonic magnons versus fermionic spinons---is a long-sought phenomenon in quantum magnetism, analogous to the confinement--deconfinement transition in quantum chromodynamics. We report evidence for such a transition in the triangular-lattice antiferromagnet TlYbS$_2$. Antiferromagnetic order develops below $T_\mathrm{N} \approx 0.53\,\mathrm{K}$. A $c$-axis field suppresses this order, driving the system into a gapless quantum spin liquid with a spinon Fermi surface above $\mu_0 H_\mathrm{c} \approx 3\,\mathrm{T}$, evidenced by a finite residual linear term in thermal conductivity, a Pauli-like susceptibility, and a temperature-independent NMR Knight shift. Approaching $H_\mathrm{c}$ from above, the scattering rate of itinerant excitations is strongly enhanced while their density of states shows no critical enhancement, atypical of conventional magnetic quantum criticality.  These results point to a continuous confinement--deconfinement transition governed by fractionalized excitations beyond the Ginzburg--Landau paradigm.
\end{abstract}

\maketitle
\section*{INTRODUCTION}
 Quantum spin liquids (QSLs) are unconventional states of matter that appear in various insulating quantum magnets when magnetic order is avoided by strong frustrations. They are characterized by fractionalized excitations arising from strong quantum entanglement, whose understanding lies at the center of modern condensed matter physics~\cite{savary2016quantum,zhou2017quantum,knolle2019field,matsuda2025kitaev,anderson1973resonating}. Well-known examples are found in Kitaev QSLs in the honeycomb lattice, in which electron spins are decoupled into Majorana fermions and emergent $\mathbb{Z}_2$ gauge fluxes~\cite{kitaev2006anyons}.  In triangular- and kagome-lattice QSLs, the elementary excitation is a spinon, another distinct type of fermionic quasiparticle that carries spin $S$=1/2 with zero charge. For certain U(1)-gauge QSLs, such spinons can form a so-called spinon Fermi surface~\cite{lee2005u1,motrunich2005variational,lee2006doping}, analogous to the Fermi surface in a metal despite the absence of conduction electrons.

 In contrast to QSLs, conventional magnets with ordered magnetic moments support bosonic excitations known as magnons ($S$=1). Theoretical studies reveal that a bosonic magnon can be viewed as a bound state of two fermionic spinons~\cite{faddeev1981spin,ghioldi2018dynamical,ferrari2019dynamical}. The transition between bound to unbound spinons can be understood as a confinement--deconfinement  (CD) transition~\cite{read1989valence,wen1991mean}, a conceptual transition originally proposed in quantum chromodynamics (QCD) in which confined quarks (hadrons) deconfine into a quark-gluon plasma~\cite{polyakov2018gauge,greensite2011introduction}. QSLs may provide a unique opportunity to investigate such a CD transition in the solid state~\cite{mandal2011confinement, senthil2001fractionalization, sachdev2003colloquium}, yet its experimental observation and clear characterization remains elusive especially for the case of a continuous transition. In addition, the associated quantum critical behavior, if any, remains poorly understood. 

\begin{figure*}[t]
\centering
\includegraphics[width=0.9\linewidth]{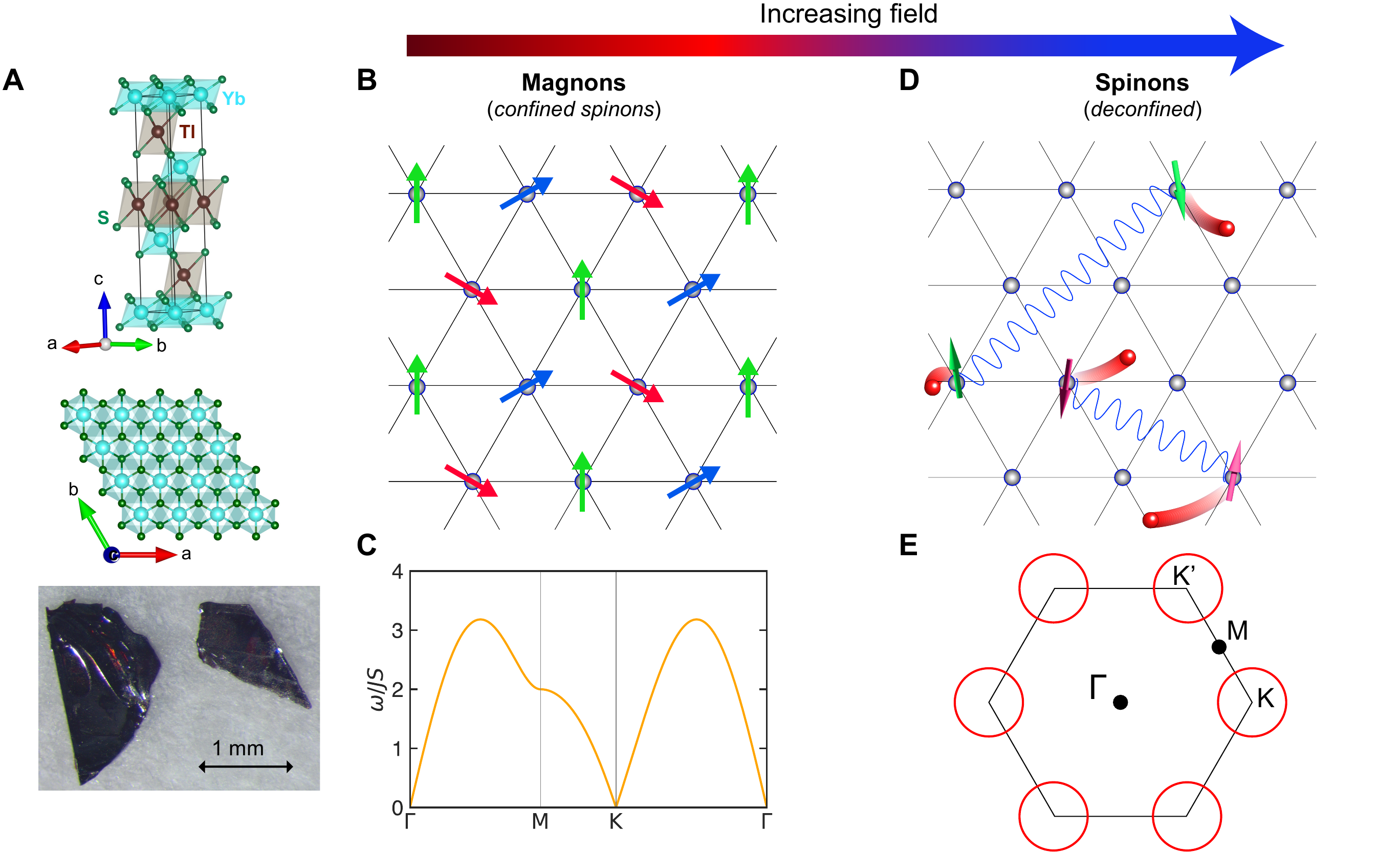}
\caption{\textbf{Crystal structure, magnetic order and quantum spin liquid state in TlYbS$_2$.}
(\textbf{A}) Trigonal crystal structure of TlYbS$_2$. Top, layered structure viewed along the $c$ axis showing Yb (cyan), S (green), and Tl (brown) atoms. Center, triangular lattice of Yb$^{3+}$ ions in the $ab$ plane. Bottom, optical image of representative thin plate-like single crystals. See Supplementary Materials for details of crystal growth and structural characterization. Crystal structures were visualized using VESTA software~\cite{momma2011vesta}.
(\textbf{B} to \textbf{E}) Schematic illustration of the confinement--deconfinement transition in TlYbS$_2$ under magnetic field applied along the $c$ axis. The field drives a continuous transition from a phase with magnon (confined spinons) excitations to one with deconfined spinon excitations.
(\textbf{B}) Schematic 120$^\circ$ noncollinear AFM order on the triangular lattice at low fields.
(\textbf{C}) Schematic magnon dispersion in the ordered phase. Solid lines indicate the gapless Heisenberg limit with Goldstone modes at the $\Gamma$ and $K$ points along the high-symmetry path $\Gamma$--$M$--$K$--$\Gamma$.
(\textbf{D}) Schematic illustration of a U(1) quantum spin liquid (QSL) with a spinon Fermi surface. Spinons (green and red arrows) are deconfined $S = 1/2$ excitations interacting via an emergent U(1) gauge field (blue wavy lines) in a quantum-entangled background (grey sites), schematically represented by red shaded regions.
(\textbf{E}) Schematic spinon Fermi surfaces in the first Brillouin zone of the triangular lattice. Red circles schematically denote the Fermi pockets of spinons.}
\end{figure*}

The $S$=1/2 two-dimensional (2D) triangular-lattice Heisenberg antiferromagnet is a prototypical frustrated quantum magnet~\cite{bernu1992signature,capriotti1999long}. Although the nearest-neighbour model stabilizes 120$^\circ$ magnetic order despite strong geometric frustration, additional interactions can destabilize the order and potentially give rise to QSL phases~\cite{zhu2015spin,hu2015competing,iqbal2016spin}. Yb-based delafossite compounds, which host an undistorted triangular lattice of effective spin-1/2 Yb$^{3+}$ ions with well-isolated 2D layers, provide an ideal platform to explore this physics~\cite{baenitz2018naybs2,ranjith2019anisotropic}. 
Several compounds in this family, including KYbSe$_2$, NaYbSe$_2$, CsYbSe$_2$, and more recently, TlYbSe$_2$, have been extensively studied~\cite{xie2023complete,dai2021spinon,Lee_2024_PhaseDiagram,scheie2024proximate,scheie2024spectrum,belbase2026finite,fujii2025tlybse}, yet the nature of their ground states remains controversial, with conflicting reports on the presence or absence of magnetic order. Whether a genuine QSL state is realized—and, if so,
the nature of its fractionalized excitations—remains an open question. In this context, Tl-based compounds are of particular interest, as their large ionic radius can provide a favorable platform for realizing a QSL state \cite{belbase2026finite}. For the sulfide analogue TlYbS$_2$, earlier bulk-susceptibility measurements found no long-range order down to 0.4\,K, suggesting a QSL candidate~\cite{ferreira2020frustrated}.

Magnetic fields have recently been used to explore putative QSL phases in several frustrated magnets~\cite{kasahara2018majorana,banerjee2018excitations,zheng2025unconventional}. Here we report evidence for a field-induced continuous CD transition in the triangular-lattice antiferromagnet TlYbS$_2$ (Fig.\,1A), between an antiferromagnetic (AFM) phase hosting bosonic magnons (Fig.\,1, B and C) and a QSL phase with fractionalized fermionic spinons.  A magnetic field ${\bm H} \parallel c$ suppresses the zero-field AFM order and drives the system into a QSL above a critical field $\mu_0H_c\approx 3$\,T. Thermal conductivity, susceptibility, and nuclear magnetic resonance (NMR) measurements consistently indicate a gapless U(1) QSL with a spinon Fermi surface in the high-field phase (Fig.\,1, D and E). The evolution across $H_c$ is inconsistent with conventional magnetic quantum criticality and instead points to a continuous CD transition between magnons and spinons.

\section*{RESULTS}
Magnetization measurements up to 60\,T (fig.\,S1A to C) reveal a sizable $c$-axis Van Vleck susceptibility, $\chi_{\mathrm{VV}}^{c} \approx 0.012\,\mathrm{emu/mol}$, which is subtracted below. The anisotropy of the saturation magnetization is in close quantitative agreement with the inverse anisotropy of the saturation field (see Supplementary Materials). A $1/3$ magnetization plateau is observed for ${\bm H}\parallel ab$, whereas it is absent for ${\bm H}\parallel c$.  Taken together, these results suggest that the spin Hamiltonian is close to the Heisenberg limit with weak easy-plane anisotropy. A comprehensive understanding of the magnetic anisotropy, including the respective roles of the $g$-factor anisotropy and exchange anisotropy, is beyond the scope of the present work and will be reported elsewhere.

\begin{figure}[!t]
\includegraphics[width=0.95\columnwidth]{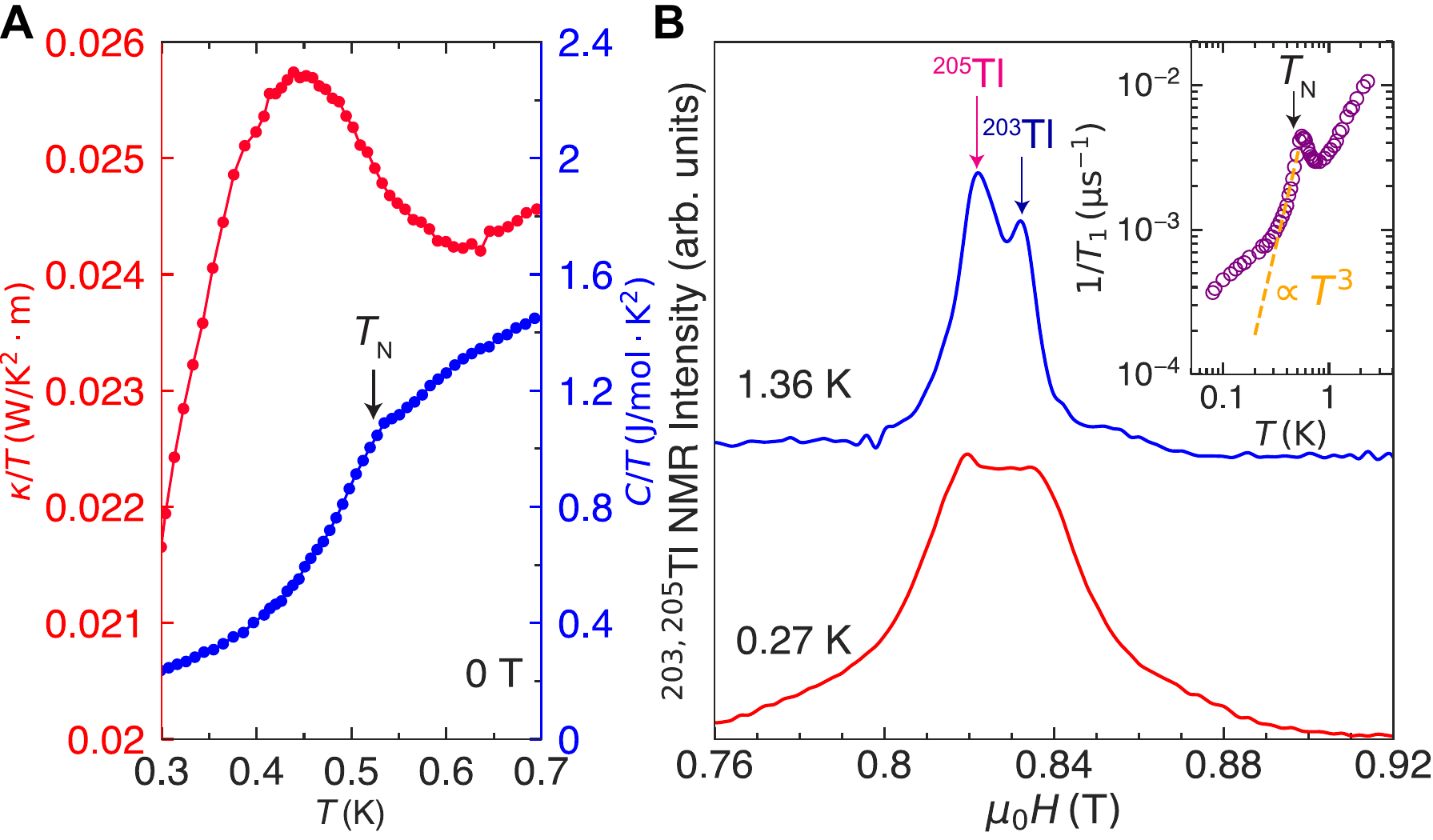}
\caption{\textbf{Evidence for AFM ordering.}
(\textbf{A}) Temperature dependence of $\kappa/T$ (red) and $C/T$ (blue) at zero field. A kink in $C/T$ and an inflection in $\kappa/T$ at $T_\mathrm{N}\approx0.53$\,K identify the onset of AFM order. On cooling, $\kappa/T$ rises below $\sim$0.6\,K and peaks at $\sim$0.45\,K as magnetic scattering is suppressed by developing spin correlations, while the broad feature in $C/T$ above $T_\mathrm{N}$ reflects the short-range correlations expected for a triangular-lattice antiferromagnet.
(\textbf{B}) $^{203,205}$Tl NMR spectra in the low-$H$ regime for $\bm{H}\parallel c$. Two sharp peaks corresponding to the $^{203}$Tl and $^{205}$Tl resonances are observed. The spectrum exhibits pronounced broadening and a substantial change in lineshape below $T_\mathrm{N}$, consistent with a growing static internal field associated with AFM order. Inset: nuclear spin-lattice relaxation rate, $1/T_1$, versus temperature at $\approx$0.83\,T. A peak at $T_\mathrm{N}\approx0.53$\,K reflects critical slowing down at the transition. The dashed line indicates a $T^3$ dependence below $T_\mathrm{N}$, characteristic of two-magnon relaxation in an ordered antiferromagnet.}
\label{fig:AFM}
\end{figure}

Figure\,2A shows the temperature ($T$) dependence of the specific heat divided by $T$, $C/T$, and the in-plane thermal conductivity divided by $T$, $\kappa/T$. At $T_{\rm N}\approx0.53$\,K, $C/T$ exhibits a kink marking the onset of magnetic order, superposed on a broad feature characteristic of short-range spin correlations in frustrated triangular-lattice antiferromagnets~\cite{capriotti1999long,schmidt2015thermodynamics}. The relatively small anomaly at $T_{\rm N}$ compared with those of conventional 3D magnets indicates that much of the magnetic entropy is released above the ordering transition, as expected for a frustrated magnet. The presence of AFM short-range spin correlations above $T_{\rm N}$ is further supported by the progressive suppression of the broad feature under magnetic field (fig.\,S2A). A similarly subtle specific-heat anomaly accompanies the neutron-confirmed 120$^\circ$ magnetic order in the related compound KYbSe$_2$~\cite{scheie2024proximate}. The thermal conductivity $\kappa/T$ increases below $\sim$0.6\,K, exhibits an inflection near $T_{\rm N}$, and reaches a maximum at $\sim$0.45\,K upon further cooling.

Below \(T_{\mathrm N}\), the \(^{203,205}\)Tl NMR spectrum at \(\sim\)0.83\,T (\({\bm H} \parallel c\)) broadens and exhibits a clear change in lineshape, rather than simply undergoing a uniform broadening, consistent with the development of static internal fields in the ordered state. The spin--lattice relaxation rate $1/T_1$ exhibits a pronounced peak at $T_{\rm N}\approx0.53$~K (Fig.\,2B, inset), characteristic of critical slowing down at a continuous phase transition. Because AFM critical fluctuations occur at finite wavevector, $1/T_1$ is generally more sensitive to the transition than the uniform ($q=0$) susceptibility. Below $T_{\rm N}$, $1/T_1$ decreases approximately as $T^3$ (Fig.\,2B, inset), consistent with two-magnon relaxation in an ordered antiferromagnet. At the lowest temperatures, $1/T_1$ crosses over to a weaker temperature dependence, a behavior discussed below. The coincidence of anomalies in $C/T$, $\kappa/T$, and $1/T_1$ at $T_{\rm N}\approx0.53$~K provides strong evidence that the transition is an intrinsic bulk property of TlYbS$_2$ rather than arising from disorder or impurity effects (see Supplementary Materials). The ordered phase is therefore most naturally attributed to a 120$^\circ$ AFM order expected for a triangular-lattice antiferromagnet (Fig.\,1B).

\begin{figure}
\includegraphics[width=0.95\columnwidth]{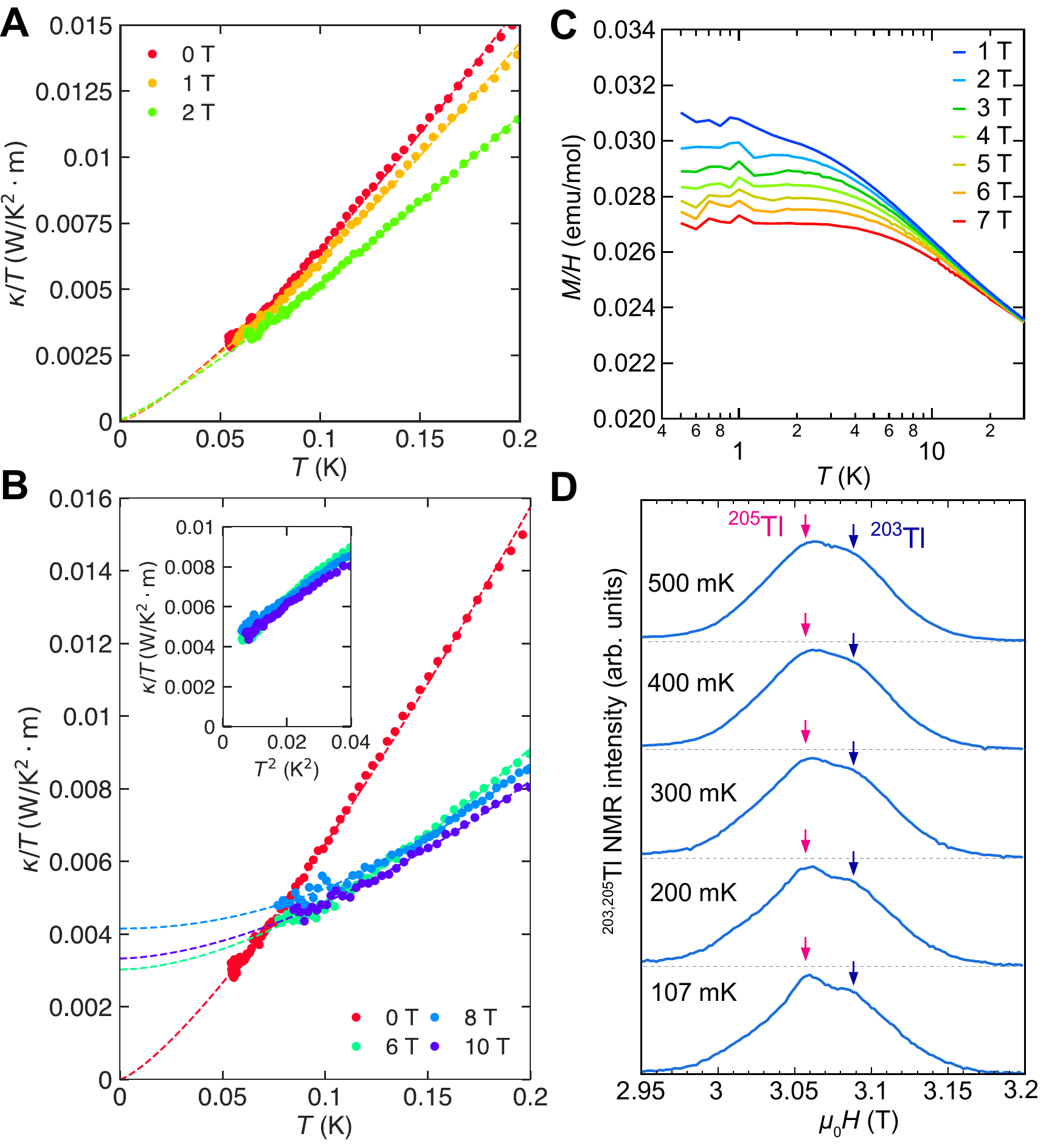}
\caption{\textbf{Evidence for a field-induced U(1) quantum spin liquid state with a spinon Fermi surface.}
(\textbf{A}) Temperature dependence of $\kappa/T$, at low magnetic fields for $\bm{H}\parallel c$. The dashed lines represent polynomial fits to the data, used to estimate the $T\rightarrow0$ limit. The extrapolated residual term $\kappa_0/T$ vanishes in the AFM ordered state.
(\textbf{B}) $\kappa/T$ at higher magnetic fields in the high-field phase. A finite residual term $\kappa_0/T$ emerges, indicating the appearance of mobile gapless excitations. Inset: $\kappa/T$ plotted against $T^2$, demonstrating the finite residual term $\kappa_0/T$.
(\textbf{C}) Temperature dependence of the magnetic susceptibility $M/H$ measured at several magnetic fields for $\bm{H}\parallel c$. A $T$-independent Pauli-like contribution develops at low temperatures, indicating a finite low-energy density of states.
(\textbf{D}) Low-$T$ $^{203,205}$Tl NMR spectra measured in the high-field phase for $\bm{H}\parallel c$. No clear spectral modulation or splitting is observed, indicating the absence of static long-range order. The red and blue arrows indicate the peak positions of the $^{205}$Tl and $^{203}$Tl spectra, respectively. They remain unchanged down to 107\,mK, indicating a $T$-independent Knight shift.}
\label{fig:spinon}
\end{figure}

\begin{figure*}[!t]
\includegraphics[width=1.2\columnwidth]{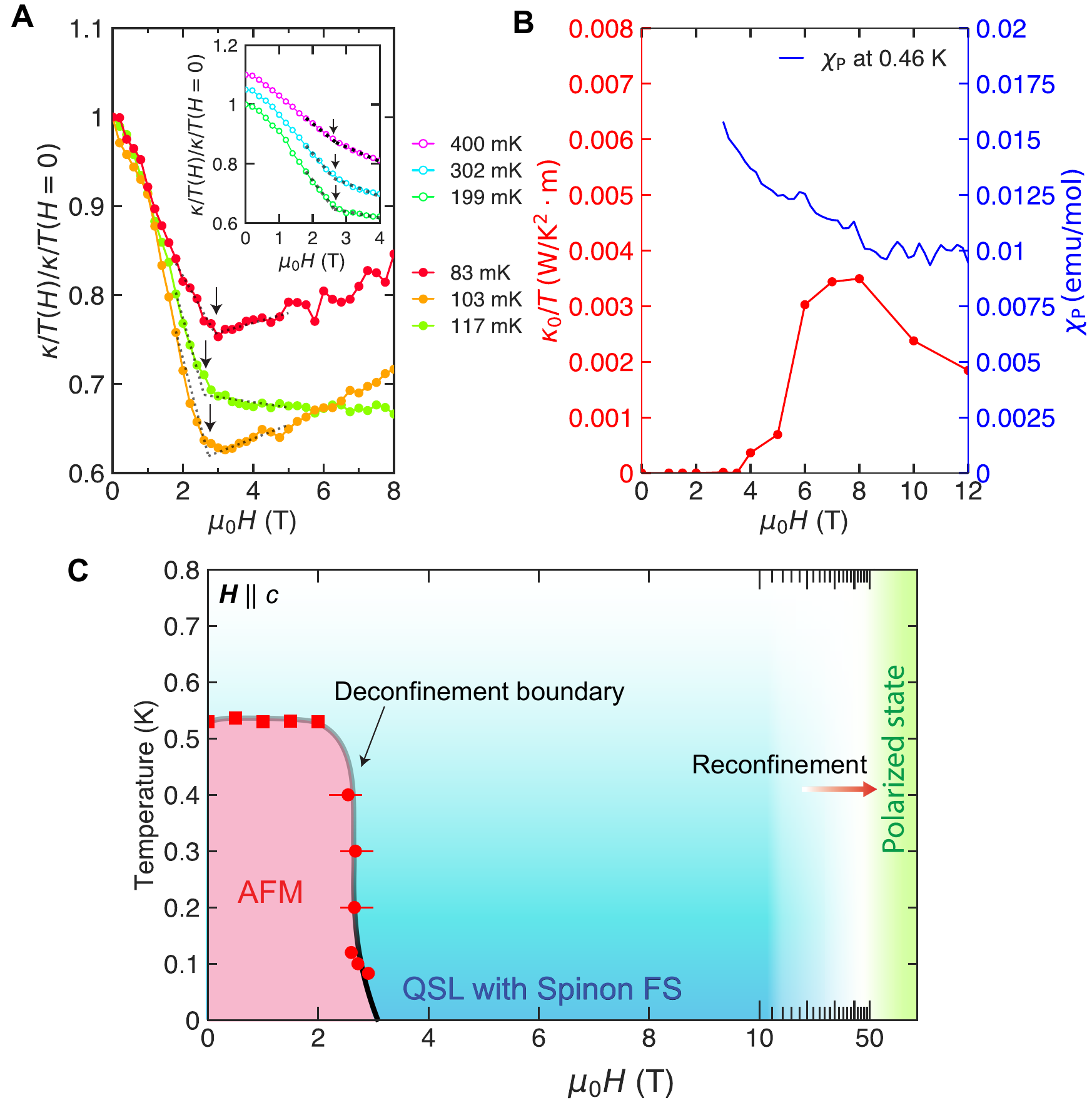}
\caption{\textbf{$H$--$T$ phase diagram of TlYbS$_2$.}
(\textbf{A}) Magnetic-field dependence of $\kappa/T$ for $\bm{H}\parallel c$, normalized by its zero-field value, measured at several temperatures. A clear kink is observed at $\mu_0 H\approx2.92$\,T at 83\,mK, consistent with a continuous phase transition between the AFM and QSL phases. Arrows indicate the kink positions used to determine the phase boundary. Inset: field dependence of $\kappa/T$ at higher temperatures, where the anomaly broadens but remains visible. For clarity, the data are vertically offset by 0.05.
(\textbf{B}) Field dependence of the residual linear term $\kappa_0/T$ obtained from low-$T$ extrapolation of $\kappa/T$, together with the Pauli-like susceptibility $\chi_\mathrm{P}$ measured at 0.46\,K in pulsed field for $\bm{H}\parallel c$. The residual linear term remains negligible at low fields and becomes finite above $\sim$3\,T. Whereas $\chi_\mathrm{P}$ evolves smoothly without a divergent enhancement, $\kappa_0/T$ is continuously suppressed toward zero on approaching $H_{\rm c}$ from the QSL side, consistent with zero-temperature criticality associated with the transition.
(\textbf{C}) $H$--$T$ phase diagram of TlYbS$_2$ for $\bm{H}\parallel c$ determined from the present measurements. The AFM phase is suppressed at $\mu_0 H_{\rm c}\sim3$\,T, above which a gapless U(1) QSL with a spinon Fermi surface persists over a wide field range. Transition points are determined from the kink in $C/T(T)$ at higher $T$ (red squares; fig.\,S2A) and from the kink in $\kappa(H)$ at lower $T$ (red circles). The observation of Pauli-like susceptibility above $T_{\rm N}$ at lower fields suggests that the AFM phase emerges from a strongly fluctuating background and remains in close proximity to the field-induced QSL. A possible high-$H$ reconfinement regime is indicated, although its nature remains unresolved. The horizontal axis uses a linear (logarithmic) scale below (above) 10\,T to cover the full field range toward the polarized state above $\mu_0 H\sim48$\,T.}
\label{fig:HTphase}
\end{figure*}

The kink in $C/T$ is barely discernible at 2\,T and absent at 3\,T (fig.\,S2B) and above. In addition, the peak of $\kappa/T$ is strongly suppressed by a $c$-axis magnetic field of 2\,T (fig.\,S2C). These results indicate that the magnetic order is suppressed by a modest magnetic field. Figure~3A shows the low-$T$ part of $\kappa/T$ measured down to 70\,mK at zero field and in a magnetic field (${\bm H}\parallel c$) up to 2\,T. Extrapolation of the lowest-temperature data to $T=0$ yields a negligibly small residual linear term, $\kappa_0/T \equiv \lim_{T\to0}\kappa/T$, indicating the absence of gapless heat-carrying fermioninc excitations. This is consistent with the AFM ordered state established below $T_{\rm N}$ (Fig.\,2, A and B), because conventional bosonic excitations such as phonons and magnons do not produce a finite residual linear term as $T\to0$,  where their population vanishes according to Bose--Einstein statistics.

In striking contrast, $\kappa/T$ measured at higher fields (Fig.\,3B) exhibits qualitatively different behavior: a finite residual linear term, $\kappa_0/T$, emerges, as clearly highlighted by the finite intercept in the $\kappa/T$ versus $T^2$ plot (inset of Fig.\,3B). The high-field thermal conductivity cannot be described by a single power law over the entire measured temperature range (fig.\,S3, A to D). The emergence of a finite $\kappa_0/T$ demonstrates heat transport by itinerant gapless spin excitations and is consistent with a transition from a ``spinon-insulating'' state to a ``spinon-metallic'' state with a spinon Fermi surface. To exclude possible effects of sample inhomogeneity, cracks, and geometric uncertainties in the thermal-conductivity measurements, we measured a second crystal from a different growth batch with a thermometer-contact spacing approximately one-tenth that of the primary sample. The absence of a residual term at zero field and its emergence in the high-field phase were reproduced (fig.\,S5), confirming that the observed thermal-transport behavior is intrinsic.

Support for this interpretation also comes from the magnetic susceptibility, $M/H$ (${\bm H}\parallel c$, Fig.\,3C), which exhibits a temperature-independent, Pauli-like behavior down to 500\,mK, consistent with a finite low-energy density of states. In the present compound, the $M$--$H$ curve is nearly perfectly linear above 0.5\,T, indicating a negligibly small impurity contribution (fig.\,S4).

Figure\,3D depicts $^{203,205}$Tl NMR spectra for ${\bm H}\parallel c$ at 3.08\,T  at low temperature. In contrast to the lower-field results (Fig.\,2B), no notable line broadening or modulation appears in either the $^{203}$Tl or $^{205}$Tl NMR spectra, and a well-defined paramagnet-like lineshape persists down to 107\,mK. The robust lineshape provides strong evidence for the absence of static long-range magnetic order in the high-field phase.

The spectral peak positions, which yield the Knight shift as the sum of spin and orbital contributions, $K = K_{\rm spin}+K_{\rm orb}$, remain unchanged below $\sim$1\,K. For $^{205}$Tl, the measured shift is $^{205}K \approx 1.8\%$. Subtracting the orbital contribution estimated from higher-temperature measurements, $^{205}K_{\rm orb}\approx -1.2\%$ (fig.\,S6), yields a finite and temperature-independent spin component, $^{205}K_{\rm spin}\approx 3.0\%$, which directly probes the local spin susceptibility. Essentially identical results are obtained at 5.65\,T, deep in the high-field phase (fig.\,S7). The persistence of a finite and temperature-independent Knight shift demonstrates that the local spin susceptibility remains Pauli-like throughout the high-field phase.

Such a finite density of low-energy states would be expected to produce a linear-in-$T$ contribution to the specific heat. Although the specific heat measured down to 50\,mK is dominated by a large nuclear Schottky contribution from Yb nuclei (fig.\,S8A), analysis of  the low-temperature data reveals a discernible linear-in-$T$ contribution at 3\,T and above in the field-induced QSL phase (fig.\,S8B). This observation provides additional support for the presence of low-energy excitations in the QSL phase. In the AFM phase at zero field, by contrast, a reliable estimate of the linear term cannot be obtained, because magnon contributions are not captured by the fitting procedure and the apparent intercept may be an artifact (see Supplementary Materials).

Taken together, the absence of static magnetic order, the finite residual $\kappa_0/T$, the Pauli-like susceptibility, the temperature-independent Knight shift, and the discernible linear-in-$T$ contribution to the specific heat at 3\,T and above provide compelling bulk evidence for a gapless QSL with a finite density of low-energy excitations. These observations are naturally explained by fermionic spinons forming a spinon Fermi surface.

Before proceeding with this exotic possibility, it is worth considering alternative scenarios. One possibility is a nearly gapless bosonic ``moat-band'' excitation spectrum, in which a ring-like dispersion minimum produces an enhanced low-energy density of states that compensates for the thermal suppression of bosonic occupation~\cite{knolle2017excitons}. However, such a scenario would require fine tuning over the entire field range and does not naturally account for the observed Pauli-like susceptibility, temperature-independent Knight shift, and specific-heat data suggestive of a finite linear term  in the high-field phase.

Two disorder-driven scenarios must also be considered. The first is conventional spin-glass freezing, which is excluded because the high-field NMR line shows no broadening down to 107\,mK---i.e., no static internal
fields develop---while both the bulk susceptibility and the Knight shift remain temperature independent, all inconsistent with frozen moments. The second, more subtle, possibility is a random-singlet or valence-bond-glass state, which does not freeze and can in principle reproduce a Pauli-like bulk susceptibility and a linear-in-$T$ specific heat~\cite{kimchi2018valence,ramirez2025short}. However, this scenario is incompatible with our data. Most importantly, the low-energy excitations in a random-singlet state are spatially localized and therefore cannot contribute to heat transport, in contrast to the finite residual $\kappa_0/T$, which requires itinerant excitations. Moreover, although a random-singlet state may mimic a Pauli-like bulk susceptibility, it generally produces a broad distribution of local susceptibilities and a pronounced temperature dependence of the local magnetic response upon cooling, inconsistent with the nearly temperature-independent Knight shift observed down to 107\,mK. These considerations rule out a random-singlet or valence-bond-glass scenario and instead support a gapless QSL hosting itinerant low-energy excitations.

To further reveal the nature of the transition between the low-field AFM and high-field QSL states, we examine the $c$-axis $H$ dependence of $\kappa/T$ at low $T$ normalized to the zero-field value (Fig.\,4A). At lower fields, $\kappa/T$ decreases with $H$, consistent with enhanced spin--phonon scattering from low-energy magnetic fluctuations that proliferate as $H$ approaches $H_{\rm c}$. At finite temperatures, where the thermal conductivity is still dominated by phonons, itinerant spin excitations in the high-field phase enhance spin–phonon scattering, thereby suppressing the phonon thermal conductivity. This finite-temperature suppression should be distinguished from the residual thermal conductivity in the zero-temperature limit. While the former is governed by spin–phonon scattering, the latter directly reflects heat transport by itinerant gapless spin excitations because the phonon contribution vanishes as $T\rightarrow 0$.

At higher fields, however, the field dependence changes qualitatively. At 83\,mK, $\kappa(H)/\kappa(H=0)$ exhibits a distinct kink at $\mu_0H \approx 2.92\,\mathrm{T}$, above which it increases nearly linearly with $H$. This kink identifies a critical field of $\mu_0H_{\rm c}\approx 3\,\mathrm{T}$ separating the AFM and QSL phases. The kink feature becomes progressively smeared out with increasing $T$, but a weak remnant remains visible up to $\sim400$\,mK (inset of Fig.\,4A). As shown in fig.\,S9, field-up and field-down sweeps coincide within experimental uncertainty, indicating no detectable hysteresis near the transition. The distinct kink anomaly, together with the absence of hysteresis, is consistent with a continuous transition and disfavors a first-order transition. Moreover, $\kappa_0/T$ decreases continuously toward zero as $H$ approaches $H_{\rm c}$ from the QSL side, suggesting proximity to a continuous quantum phase transition associated with the field-induced transition.

In Fig.\,4B, the Pauli-like temperature-independent susceptibility $\chi_{\rm P}$ ($={\rm d}M/{\rm d}H-\chi_{\rm VV}^{\rm c}$), measured at 0.46\,K, is presented for fields above $H_{\rm c}$ up to 12\,T. The nearly temperature-independent Knight shift in the QSL state down to the lowest temperatures suggests that $\chi_{\rm P}$ in Fig.\,4B closely reflects its zero-temperature value. Over the entire field range above $H_{\rm c}$, $\chi_{\rm P}$ remains finite, indicating that the low-energy density of states remains intact throughout the high-field phase. Notably, while $\chi_{\rm P}$ is largest near $H_{\rm c}$, it shows no evidence of critical divergence and decreases monotonically with increasing field.

Remarkably, the Pauli-like temperature-independent susceptibility persists to fields below 3\,T in the paramagnetic regime above $T_{\rm N}$ (Fig.\,3C). The AFM phase remains stable up to 2\,T with nearly unchanged $T_{\rm N}\approx0.53$\,K (fig.\,S2B), indicating that the Pauli-like response survives just above the AFM ordered phase. Evidence for persistent low-energy fluctuations also comes from the nuclear spin dynamics: as noted earlier, $1/T_1$ in the AFM state crosses over to a weaker temperature dependence in the low-temperature regime (inset of Fig.\,2B), indicating that low-energy fluctuations persist within the ordered state.  Taken together, these observations suggest that the AFM phase emerges from a strongly fluctuating background and remains in close proximity to the field-induced U(1) QSL.

The resulting picture is summarized in the $H$--$T$ phase diagram shown in Fig.\,4C for $\bm{H}\parallel c$. The AFM phase is suppressed at $H_{\rm c}$, above which a gapless U(1) QSL phase with a spinon Fermi surface emerges. The field-induced emergence of such a QSL phase raises the question of its microscopic origin. One possible scenario is that the zero-field ordered state lies in close proximity to the widely discussed U(1) Dirac QSL state. In this case, a magnetic field may induce an effective spin scalar chirality, which has been argued to stabilize a gapped chiral QSL phase over an extended field range~\cite{cookmeyer2021four,yang2026emergent}. Fermi pockets near the Brillouin-zone corners could then emerge from this chiral phase through mechanisms such as Zeeman splitting~\cite{yang2026emergent} or inversion-symmetry breaking~\cite{chari2021magnetoelectric}. Whether such a scenario is realized in TlYbS$_2$ remains an open question and will require further theoretical and experimental investigation.

\section*{Discussion}
The phase transition from the AFM phase to the gapless QSL phase in TlYbS$_2$ is analogous to a CD transition discussed in QCD in the context of quark confinement. In contrast to the generally first-order or crossover-like behavior expected in QCD, however, the transition observed in our system is continuous. It may alternatively be viewed as a metal--insulator transition of emergent neutral spinons, from a ``spinon thermal insulator'' in the confined antiferromagnet to a ``spinon thermal metal'' in the deconfined QSL, even though TlYbS$_2$ remains an electrical insulator throughout. Here, $\kappa_0/T$ serves as the thermal analogue of the electrical conductivity in a conventional metal--insulator transition, providing a transport measure of the deconfinement transition and vanishing continuously at $H_{\rm c}$. Within this framework, the bosonic magnons of the AFM phase may be viewed as confined spinons, whereas the gapless QSL hosts deconfined spinons coupled to an emergent U(1) gauge field~\cite{senthil2008theory} (Fig.\,1, B to E). The transition therefore lies beyond the conventional Ginzburg--Landau framework, involving a fundamental change in the nature of the low-energy quasiparticles. A comparison with $\alpha$-RuCl$_3$ is instructive: an in-plane field there suppresses AFM order and is proposed to induce a gapped $\mathbb{Z}_2$ QSL hosting fractionalized Majorana fermions~\cite{banerjee2016proximate,kasahara2018majorana}, but that transition is reported to be first-order~\cite{schonemann2020thermal} and the Majorana gap does not close at the transition field~\cite{matsuda2025kitaev}. By contrast, the continuous quantum phase transition in TlYbS$_2$ gives rise to a gapless spinon Fermi-surface state beyond $H_{\rm c}$, providing a rare platform for studying a continuous confinement--deconfinement transition between bosonic magnons and fermionic spinons.

The modest enhancement of $\chi_{\rm P}$ upon approaching $H_{\rm c}$ from the QSL side is reminiscent of behavior often found near a magnetic quantum critical point (QCP). However, the susceptibility remains essentially temperature-independent and Pauli-like down to 107\,mK even in the vicinity of $H_{\rm c}$, as evidenced by the NMR Knight shift (Fig.\,3D), showing neither the critical divergence nor the singular temperature dependence generally expected near a conventional magnetic QCP. These observations are difficult to reconcile with a conventional magnetic QCP.

Figure\,4B compares the field dependence of $\kappa_0/T$ and $\chi_{\rm P}$. As $H_{\rm c}$ is approached from the QSL side, $\kappa_0/T$ is strongly suppressed, whereas $\chi_{\rm P}$ exhibits only a modest increase. These contrasting trends suggest that spinon scattering is selectively enhanced in the vicinity of the CD transition. The $\kappa/T$ data at 83 and 103\,mK show the same tendency, decreasing approximately linearly on approaching $H_{\rm c}$ from the QSL side (Fig.\,4A), indicating that the suppression of thermal transport extends continuously to finite temperatures.

In theoretical descriptions of a QSL with a spinon Fermi surface, gauge-field fluctuations are expected to produce singular scattering of spinons, strongly affecting transport properties that are sensitive to the scattering rate, while thermodynamic quantities governed primarily by the density of states are expected to be less affected~\cite{lee1992gauge,lee2006doping}. The contrasting field dependence of $\kappa_0/T$ and $\chi_{\rm P}$ is qualitatively consistent with this scenario, suggesting that critical fluctuations of the emergent U(1) gauge field may play an important role in the quantum critical behavior of TlYbS$_2$. While it is tempting to interpret these results in terms of an unconventional QCP associated with a CD transition, its critical properties remain to be established through quantitative determination of the critical exponents and scaling behavior.

Although lying at the heart of QCD, the extreme conditions required to realize a CD transition in quark--gluon matter---temperatures of order $10^{12}\,\mathrm{K}$---have hindered detailed experimental studies of the critical properties of such transitions. By contrast, the field-induced CD transition in TlYbS$_2$ occurs under readily accessible laboratory conditions, providing a promising platform for investigating the critical behavior associated with CD transitions in a quantum material.

The magnetization data show a tendency toward saturation near $\mu_0 H$$\sim$48\,T, without any clear anomaly (fig.\,S1C). These observations, together with the phase diagram in Fig.\,4C, suggest that the field-induced QSL state with a spinon Fermi surface may persist over a rather extended field range, but may eventually be replaced at higher fields by another state in which the spinon Fermi surface disappears, possibly through a reconfinement of spinons into bosonic quasiparticles. However, magnetization measurements alone are not sufficiently sensitive to determine whether such an evolution occurs through a true phase transition or a crossover. Determining the nature of this possible high-field reconfinement and establishing a theoretical framework for the continuous CD transition beyond the Ginzburg–Landau paradigm~\cite{alet2006exotic,hermele2008properties,dupuis2019transition} remain important challenges for future studies.

\section*{Materials and Methods}
 \noindent
\textbf{Single crystal synthesis:}

Single crystals of TlYbS$_2$ were grown using a two-step method consisting of polycrystalline synthesis followed by flux growth from polycrystalline TlYbS$_2$ obtained from stoichiometric amounts of thallium granules (99.99\,\%, Thermo Fisher), ytterbium powder (99.9\,\%, Thermo Fisher), and sulfur powder (99.999\,\%, Thermo Fisher). The crystals were grown using TlCl flux at a maximum temperature of 690\,$^\circ$C leading to the formation of ruby-red, high-quality, 2D single crystals of TlYbS$_2$ with typical lateral dimensions of $\sim$2\,$\times$\,2\,mm or larger.  The crystal quality and stoichiometry were examined using single-crystal X-ray diffraction (XRD) and energy-dispersive X-ray spectroscopy (EDX). Single-crystal XRD measurements were performed at room temperature on a Bruker D8 Quest diffractometer equipped with a Mo K$\alpha$ radiation source ($\lambda = 0.71073$ \AA). Diffraction data were collected using APEX6 with $\omega$- and $\phi$-scan techniques. The crystal structure was solved and refined using the \textsc{ShelX} software package \cite{sheldrick2008short}. Single-crystal XRD measurements and refinements were performed on multiple crystals, all of which yielded consistent structural parameters and confirmed a trigonal crystal structure with space group $R\bar{3}m$ (No.~166), in contrast to the hexagonal $P6_3/mmc$ structure reported by Ferreira \textit{et al.}~\cite{ferreira2020frustrated}. No evidence of site mixing or crystallographic disorder was detected within the sensitivity of the refinement. The refined lattice parameters are $a=b=3.9299(3)$~\AA\ and $c=22.459(10)$~\AA. The refinement converged with excellent agreement factors of $R_1 = 0.0175$ and $wR_2 = 0.0374$ (all data), with a goodness-of-fit (GoF) of 1.268.

To further verify the sample composition, EDX measurements were performed on multiple single crystals. The measured elemental ratios consistently yield Tl:Yb:S $\approx$ 1:1:2, confirming the expected stoichiometry of TlYbS$_2$. Full details of the crystal growth and structural refinement will be reported elsewhere~\cite{Belbpase_preprint}.\\

\noindent
\textbf{Thermal transport:}

Thermal transport measurements were performed on two independently grown single crystals of TlYbS$_2$. The main data presented in the manuscript were obtained from sample \#1, a thin plate-shaped crystal with dimensions of $3.1 \times 0.85 \times 0.04\,\mathrm{mm}^3$. The heat current was applied within the $ab$ plane along the longest dimension of the crystal. The thermal conductivity was measured using a standard one-heater, two-thermometer steady-state technique. Gold wires were attached to the sample using silver paste to ensure good thermal contact between the sample and the thermometers. Measurements were carried out down to 70\,mK in a dilution refrigerator.

The reproducibility of the thermal-transport results was verified using an independently grown crystal (sample \#2; dimensions $1.2 \times 0.36 \times 0.03\,\mathrm{mm}^3$). Silver-wire contacts were used instead of the gold-wire contacts employed for sample \#1, and the separation between the thermometer contacts was approximately one order of magnitude shorter.\\

\noindent
\textbf{Specific heat:}

A large single crystal (1.85\,mg) was mounted on a sapphire stage, with a heater attached to the backside of the stage. A thermometer was affixed to the top surface of the sample. The specific heat was measured using a quasi-adiabatic heat-pulse method. A weak thermal link to the thermal bath was established using Au--7~$\mathrm{wt\%}$ Cu alloy wires with a diameter of 25\,\text{\textmu}$\mathrm{m}$, providing a clear separation between the external thermal relaxation time and the internal equilibration time within the stage, sample, and thermometer.\\

\noindent
\textbf{Magnetization:}

 Low-field magnetization was measured in two commercial Magnetic Property Measurement Systems (MPMS, Quantum Design), using a single crystal (2.88\,mg). An MPMS equipped with a $^3$He option was used for measurements below 1.8\,K, whereas a separate standard MPMS was used for measurements above 1.8\,K.
 
 High-field magnetization was measured using millisecond pulsed magnets up to 60\,T at the Pulsed Field Facility of the National High Magnetic Field Laboratory, Los Alamos National Laboratory. Multiple single crystals (total mass 2.03\,mg) were aligned with $\boldsymbol{H} \parallel c$, stacked, and mounted inside a nonmagnetic ampoule using Apiezon N grease to ensure good thermal contact and mechanical stability.\\

\noindent
\textbf{Nuclear Magnetic Resonance (NMR):}

Single-crystal NMR measurements were performed at $^{203}$Tl and $^{205}$Tl sites (both nuclear spin $I=1/2$; natural abundance 29.5\,$\%$ and 70.5\,$\%$; and nuclear gyromagnetic ratio $\gamma_\mathrm{n}=$ 24.327 and 24.567\,MHz/T, respectively) using the standard spin-echo pulse sequence technique with commercially available spectrometers (REDSTONE, Tecmag Inc.) and conventional helium-4 superconducting magnets equipped with a variable temperature insert cryostat or a dilution refrigerator. The trigonal crystal structure of TlYbS$_2$ (space group $R\overline{3}m$) has a unique Tl site in the unit cell that forms a triangular lattice (Wyckoff letter 3$a$ and point group symmetry $-3m$). The NMR spectrum was measured at a fixed carrier frequency by sweeping the magnetic field and integrating the Fast Fourier Transform of the spin echo signal at each field. The spin-lattice relaxation rate $1/T_1$ was determined using the inversion recovery method and by fitting the recovery of nuclear magnetization to a stretched exponential function. No detectable difference was observed in the size of  $T_1$ at $^{203}$Tl and $^{205}$Tl sites.


\section*{Acknowledgments}
 We thank S. Fujimoto, S.Z. Lin, J. Nasu, and K. Totsuka for insightful discussions. We acknowledge Piyush Challare and Eun-Sang Choi for their assistance with crystal growth and high-field in-plane magnetization measurements at the National High Magnetic Field Laboratory (NHMFL). 

 \section*{Funding}
 
 Thermodynamic and transport measurements were supported by the U.S. Department of Energy, Office of Science, National Quantum Information Science Research Centers, Quantum Science Center. NMR measurements were supported by the U.S. Department of Energy, Office of Basic Energy Sciences, Division of Materials Science and Engineering project ``Quantum Fluctuations in Narrow-Band Systems". The Purdue team acknowledges support by the U.S. Department of Energy, Office of Basic Energy Sciences, grant DE-SC0022986. A portion of this work was performed at NHMFL, which is supported by National Science Foundation Cooperative Agreement No. DMR-2128556$^*$ and the State of Florida, and the U.S. Department of Energy.
 S.H. acknowledges the Director's Postdoctoral Fellowship through the Laboratory Directed Research and Development Program at Los Alamos National Laboratory. JK acknowledges support from the Deutsche Forschungsgemeinschaft (DFG, German Research Foundation) under Germany’s Excellence Strategy–EXC– 2111–390814868, DFG grants No. KN1254/1-2, KN1254/2-1, and TRR 360 - 492547816, as well as the Munich Quantum Valley, which is supported by the Bavarian state government with funds from the Hightech Agenda Bayern Plus. 
Work in Japan was supported by a Grant-in-Aid for Transformative Research Areas (A) ``Correlation Design Science'' (No.\ JP25H01248) and by Grants-in-Aid for Scientific Research (KAKENHI) (Nos.\ JP25H00838 and JP26K21719) from the Japan Society for the Promotion of Science (JSPS) and by JST PRESTO (No.\ JPMJPR2458), Japan.

\section*{Author contributions}
S.H. and Y.M. conceived the study. B.P.B., A.U. and A.B. synthesized the samples and performed magnetization measurements. S.H. performed thermal conductivity and specific heat measurements. S. P., M.H. and A.P.D. performed NMR measurements. J.D.T., K.I. and K.H. performed dc magnetization measurements, and M.L. performed high-field magnetization measurements using a pulsed-field magnet. All authors discussed the results. S.H., M.H. and Y.M. prepared the manuscript with input from all authors.

\section*{Competing interests}

The authors declare no competing interests.

\section*{DATA, CODE, AND MATERIALS AVAILABILITY}

All data and code needed to evaluate and reproduce the results in the paper are present in the paper and/or the Supplementary Materials. Materials are available from the corresponding authors upon reasonable request.

\end{document}